# Comparative study of noise-like pulse emission from Mamyshev oscillator and conventional passively mode-locked fiber laser

Oumaima OUGRIGE, Florent BESSIN and François SANCHEZ
Laboratoire de Photonique d'Angers, EA 4464, Université d'Angers
2 Bd Lavoisier, 49000 Angers, France

## Abstract

We perform an in-depth comparison of noise-like pulse emission from Mamyshev oscillator and mode-locking through nonlinear polarization rotation (NPR) in fiber lasers. NLP's temporal and spectral characteristics are analyzed in both normal and anomalous dispersion regimes. The output coupling coefficient is modified to optimize the energy of the output pulses. It is demonstrated that the pulse energy is systematically higher in NPR-based mode-locking fiber lasers than in Mamyshev oscillators. This is due to the filtering effect in the Mamyshev oscillator which rejects ultrashort pulses whose energy is insufficient to successfully pass through the two shifted optical filters.

## 1. Introduction

Noise-like pulse emission (NLP) is a widely observed regime in passively mode-locked fiber lasers, independent of the mode-locking technique or dispersion regime [1-6]. Indeed, it has been observed using nonlinear polarization rotation (NPR), nonlinear optical loop mirror (NOLM) or nonlinear amplifying loop mirror (NALM). Noise-like pulse emission consists of a long packet, typically in the nanosecond or sub-nanosecond ranges, involving many ultrashort pulses in perpetual motion and characterized by a flat broadband optical spectrum. In the temporal domain, NLP are characterized by an increase of both the packet duration and energy versus the pumping power. NLP sources find numerous applications, for instance in supercontinuum generation or nonlinear frequency conversion [2].

Recently, a new technique for generating ultrashort pulses based on two Mamyshev regenerators has been proposed and demonstrated [7-10]. The mechanism is based on two shifted band-pass spectral filters combined with two optical amplifiers in such a way that only high energetic pulses can pass through the filters thanks to the self-phase modulation (SPM) resulting from optical Kerr effect. Mamyshev oscillator (MO) has attracted great interest due to its ability to generate high energetic pulses, in contrast to classical mode-locking techniques that are limited by wave breaking [11]. Its major drawback is that, while the cavity is blocked in linear (continuous wave regime) or weakly nonlinear regime, the system is not self-starting and requires external control to initiate the oscillation [12-15]. One of the most useful methods consists of injecting external pulses [12]. Once the oscillation starts, the seed source can be switched off, the Mamyshev oscillation maintaining itself. In a recent paper we proposed an original self-starting all-fibered Erbium-doped Mamyshev oscillator [16]. Two tunable band-pass spectral filters were employed to initiate mode-locking through the NPR mechanism when their spectral window fully overlap. The central frequencies of the filters were then progressively shifted to achieve full filters separation, while short pulse generation was maintained through the Mamyshev mechanism. Due to the starting mechanism favoring multi-pulse emission in the form of NLP, the Mamyshev operated spontaneously in noise-like pulse regime.

Many studies on Mamyshev oscillators have focused on pulse energy in single pulse regime and on its comparison with the energy achievable with conventional mode-locking techniques [17-19]. Although NLP emission has been recently reported in MO [16,20], there is no comparative studies between NLP characteristics in MO and NPR mode-locking. This is the motivation behind this work which is devoted to an in-depth comparison of NLP emission from Mamyshev oscillator and NPR based mode-locking. A direct comparison is possible because a simple adjustment of the filters allows us switching between NPR and Mamyshev mechanisms while all other experimental parameters remain unchanged. In section 2, we present the experimental setup together with the procedure to start the oscillation. Anomalous and normal dispersion regime are considered in section 3 and 4, respectively. In both sections, in order to extract a maximum pulse energy, different output coupling coefficients are tested, and the NLP regime was characterized through spectral bandwidth, pulse energy and duration.

**2. Experimental setup and starting of the Mamyshev oscillator**

The experimental setup is represented in Fig. 1. It is an all-fiber 47 meter-long ring cavity composed of two cascaded Mamyshev regenerators (MRs) [16]. Each MR incorporates an Erbium-doped fiber amplifier (EDFA) as its gain medium (A1 and A2 in Fig. 1). To ensure unidirectional light propagation and eliminate parasitic reflections, two polarization-independent isolators are integrated into the cavity, after the amplifying sections (ISO in Fig. 1). Additionally, the cavity includes a polarization-dependent isolator placed between two polarization controllers to enable the NPR technique (PD-ISO in Fig. 1). Two tunable super gaussian filters with adjustable bandwidth ($\Delta\lambda_F$) and central wavelength ($\lambda_{c,F}$) are placed in each MR arm, right before amplifiers (F1 and F2 in Fig. 1). These filters offer a tunable bandwidth ranging from 1 nm to 18 nm in the C-band, which is achieved by precisely adjusting their transmission edges. The NPR mechanism combined with the total overlapping of the two filters spectral window, is essential for initiating mode-locking in this MO configuration. A fiber coupler is used to extract the laser output for experimental measurements. As described later in this study, different output coupling ratios were tested in order to optimize the energy of the output pulses. This output signal was recorded in the slow and fast time domain using a 12 GHz-bandwidth photodiode combined with an oscilloscope (Agilent Infinium DS08134B, 13 GHz bandwidth), and an optical autocorrelator (Femtochrome FR-103XL), respectively. The optical and radio frequency spectra are recorded with an optical spectrum analyzer (Anritsu MS9740A), and a radio frequency spectrum analyzer (Rohde and Schwarz FSP (9kHz-13.6GHz) combined with a 12 GHz-bandwidth photodiode), respectively.

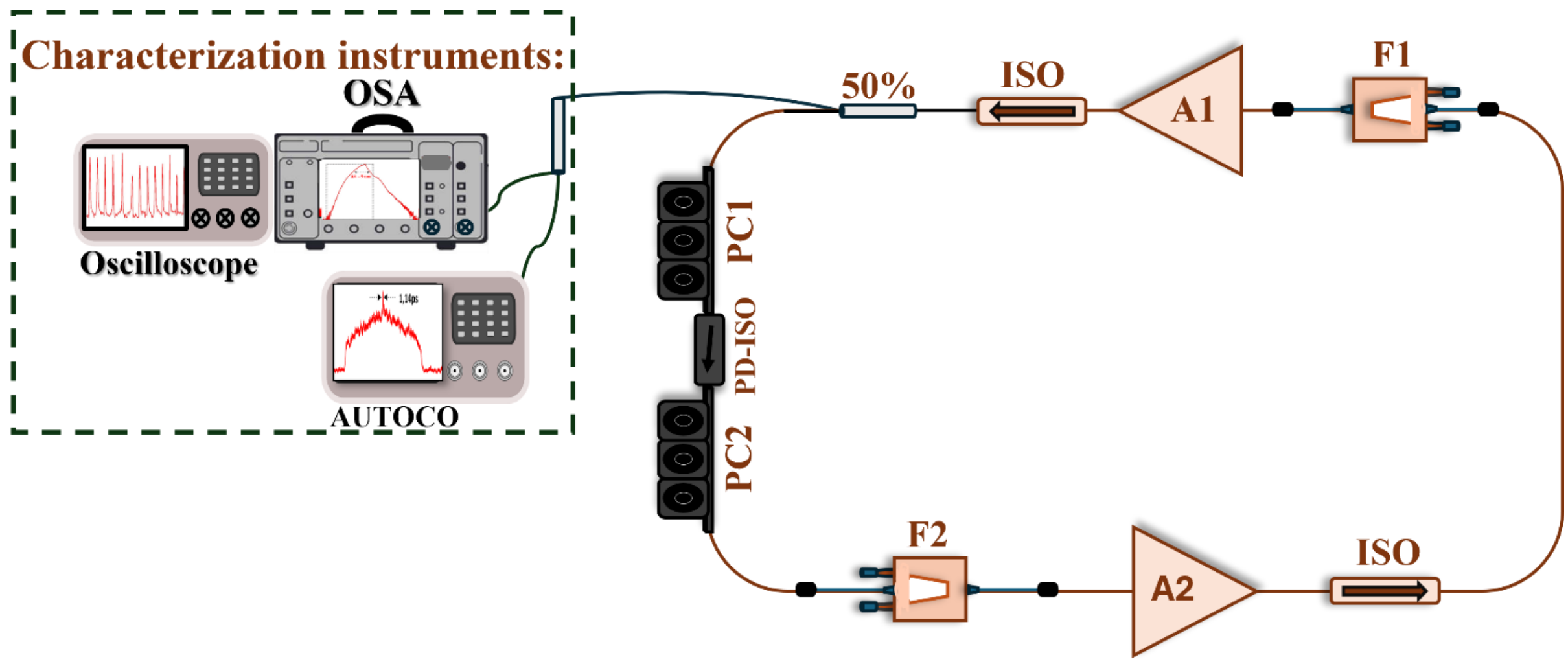

**Fig. 1**. Experimental setup.

The core principle behind mode-locking based on Mamyshev mechanism in this configuration follows three key steps [16]. First, the two filters were fully overlapped, each centered at 1550 nm with a total bandwidth of 18 nm. The pump powers of the two amplifiers were arbitrarily fixed. Through adjustment of the polarization controllers, mode-locking occurred due to the NPR mechanism. The laser operated generally in noise-like pulse regime as shown in Fig. 2(a). The resulting optical spectrum exhibits a bandwidth of approximately 9 nm at the 3 dB level, centered around a wavelength of 1555 nm. In the time domain, the signal consists of a nanosecond envelope involving many moving ultrashort pulses. Subsequently, once the noise like pulses were circulating within the cavity, we progressively began separating the central wavelengths of the two tunable filters by manipulating their upper and lower wavelength positions, respectively. This has enabled us to move from complete to partial spectral overlap. Additionally, to maintain the NLP regime under these conditions, it was necessary to increase the pump power of each amplifier accordingly. As the spectral overlap between the two filters was reduced to approximately 3 nm, the intracavity noise like pulses underwent significant spectral broadening due to SPM. This nonlinear broadening occurs in the cavity's gain fiber and single-mode fiber (SMF) segments. When the filters are partially overlapped, the resulting spectrum, shown in Fig. 2(b), is centered at 1551 nm with a 3 dB bandwidth of about 11 nm. Lastly, full spectral separation of the two filters is required to fully establish oscillation based on the Mamyshev mechanism. As illustrated in Fig. 2(c), the output spectrum exhibited a more broadened bandwidth of about 20 nm. The upper position of filter 1 and the lower position of filter 2 are separated by 1 nm, and filters bandwidths are about 11 nm and 6 nm, respectively. Furthermore, by increasing the pump power, we were able to separate both filters up to 7 nm without losing the Mamyshev oscillation regime and while still maintaining stable NLP operation. It is worth noting that following these steps did not lead to single-pulse mode-locking, instead, we consistently obtained the NLP regime based on Mamyshev mechanism. Even when pump power is reduced, the MO operates in NLP regime until it switches off.

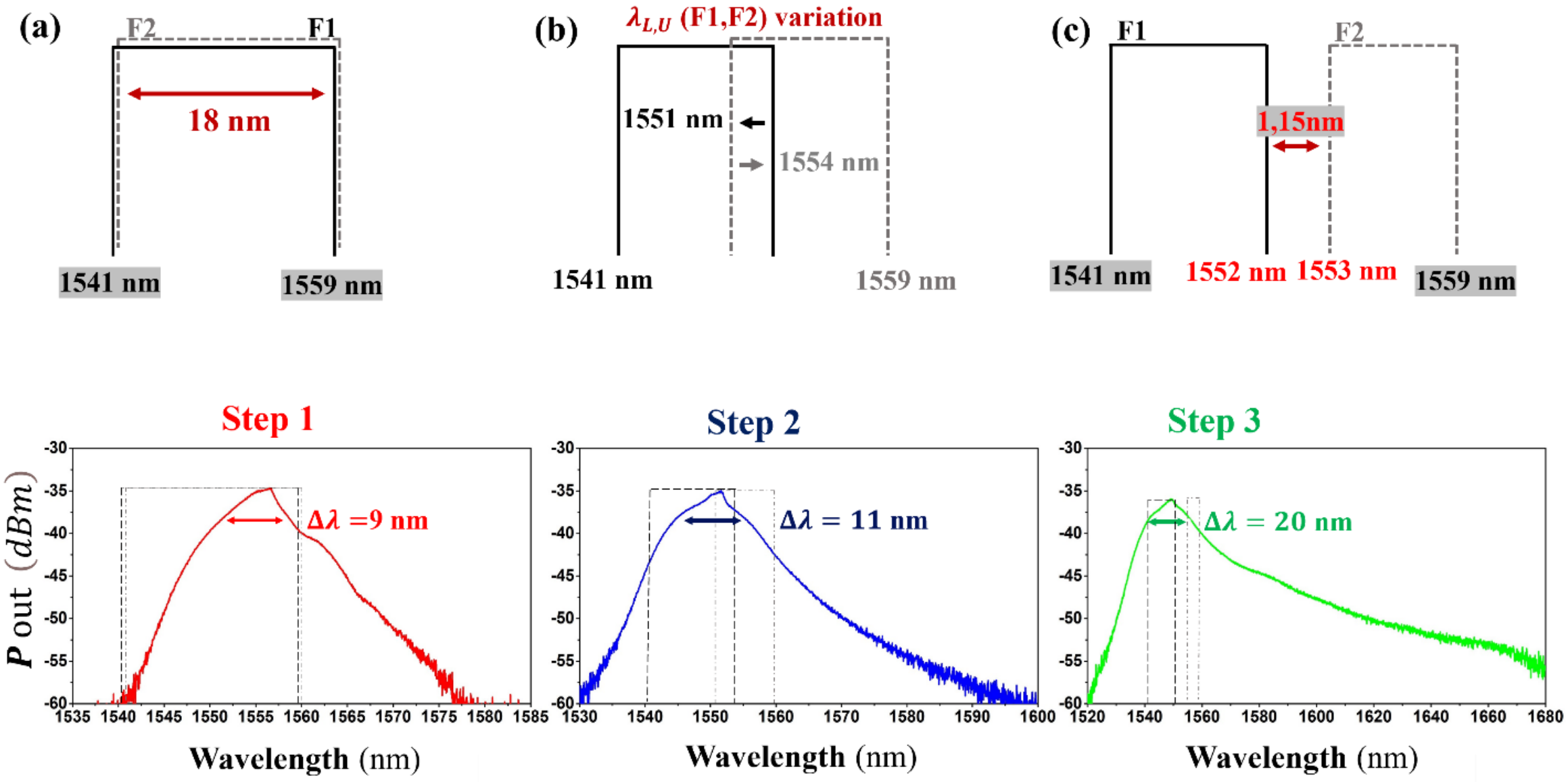

**Fig. 2**. Filters configurations and output spectra for each case. (a) overlapped filters (red spectrum), (b) partially overlapped filters (blue spectrum), (c) totally separated filters (green spectrum).

A typical example of temporal trace recorded with the photodiode and oscilloscope is shown in Fig. 3(a) in the case of NPR mode-locking mechanism. The duration of the packet can significantly change from one round-trip to the other. The corresponding autocorrelation trace is also given in Fig. 3(b). It exhibits a broad pedestal with a coherent central peak, characteristics of noise-like pulse emission [2]. Very similar results are obtained in all tested configurations.

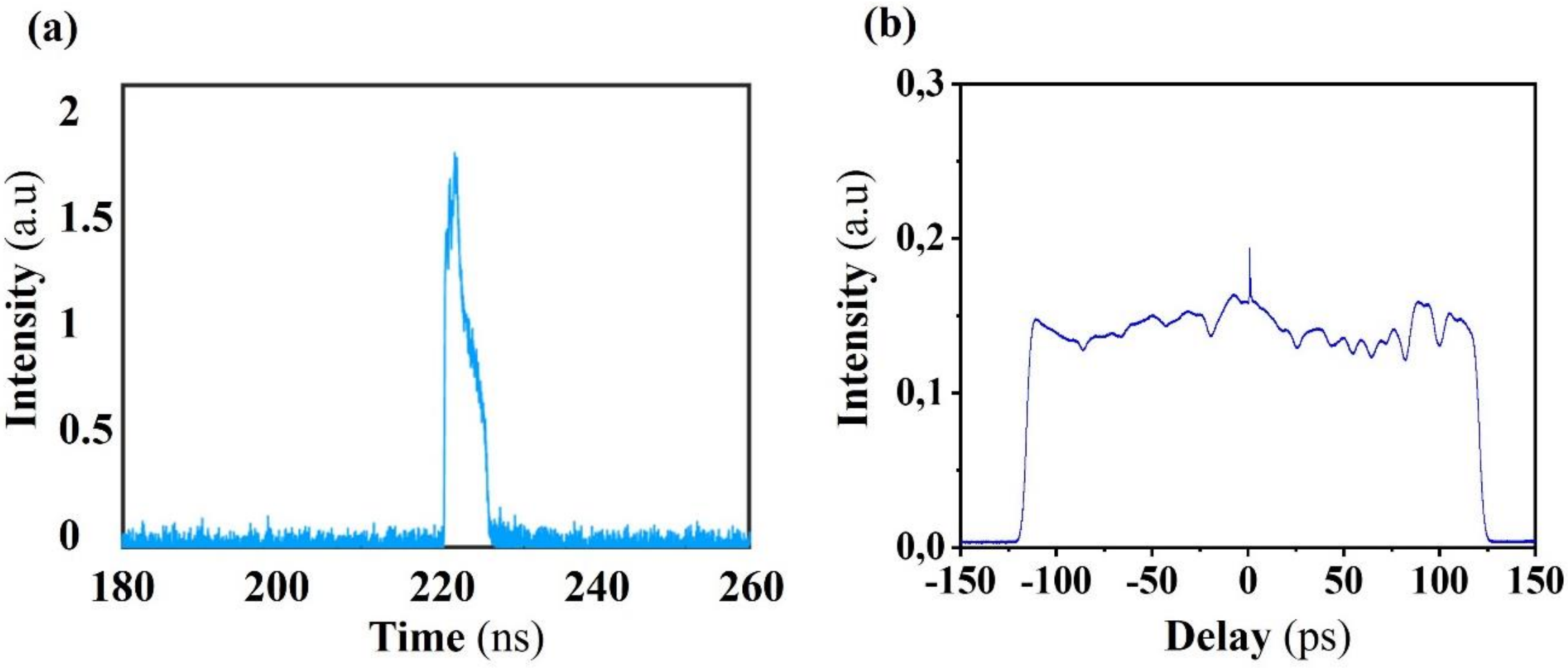

**Fig. 3.** (a) Typical temporal trace of a noise-like pulse, (b) corresponding autocorrelation signal.

### 3. Anomalous dispersion regime

Let us now compare the properties of the NLP emission from MO and NPR mode-locked laser. Such direct comparison is possible because both mode-locking techniques are realized with the same optical configuration with a simple tuning of the optical filters. In this section we consider the anomalous dispersion case for which the cavity length is 45 m and the total cavity dispersion is estimated to be $-0.94\ \text{ps}^2$. We will consider three output coupling coefficients $T = 20\%, 50\%, 80\%$. In each case, we first established the main conditions required to achieve the NLP regime in conventional mode-locking. In this configuration, the laser cavity was fully closed by completely overlapping the two spectral filters, with the lower edge fixed at 1555 nm and the upper edge at 1570 nm (for both filters), resulting in a bandwidth of 15 nm. The pump power in amplifiers A1 and A2 were initially set to approximately 2.2 W and 0.8 W, respectively, and then gradually increased to maintain the NLP regime in a conventional mode-locked laser cavity. The polarization controllers were kept fixed during this procedure. Afterwards, the spectral filters were progressively separated along with increasing the pump powers to approximately 3.5 W and 1.4 W, respectively. Under these conditions, we achieved the first NLP regime with a filter separation of about $\delta\lambda = 2$ nm, referring to the distance between the lower edge of one filter and the upper edge of the other, not between their center wavelengths. The NLP regime persisted as the separation was further increased up to approximately 7 nm. Then the pump powers were also increased progressively for each case to analyze the characteristics of the pulses. In the following, we focus on three filter separations of $\delta\lambda = 2$ nm, 4 nm and 7 nm. Each case was investigated under different pump powers for the three output coupling ratios given previously.

For each set of parameters (output coupling ratio $T$, filters separation $\delta\lambda$), NLP was characterized through the measurement of pulse energy and duration $\Delta\tau$ as a function of the pump power. Note that the other parameters, such as the polarization controllers, remain nearly unchanged during the data acquisition. Experimental results are summarized in Fig. 4. Lines a) to c) stands for experimental results obtained with output coupling coefficient $T = 20\%, 50\%, 80\%$, respectively. The left column gives the evolution of the pulse energy as a function of the pump power. It can be noticed that the energy increases linearly with the pump power and, if we compare, for instance the NPR results (green, cyan, and yellow curves in Fig. 4 a),b) and c), respectively), we observe that the energy, for a fixed pumping power, increases when the output coupling increases. This agrees with the fact that the output coupling optimization in fiber lasers is realized for high values of $T$, typically $T \geq 90\%$ [21]. For a given value of $T$, the pulses appear only at higher pump powers in the MO compared to the NPR mechanism. This can be explained by the fact that increasing the filter separation requires higher pump power to generate ultrashort pulses with sufficient energy to successfully pass through both filters. Indeed, the Mamyshev mechanism eliminates many short pulses with insufficient energy [8] whereas such pulses are not filtered with the NPR mode-locking. Consequently, the NLP energy in the MO decreases as the filter separation $\delta\lambda$ is increased due to the elimination of low energetic ultrashort pulses. In summary, the NPR configuration achieves higher energies but requires slight adjustment of the polarization controllers. Conversely, the MO configurations rely on spectral filtering, which acts as a nonlinear gate that preferentially transmits the strongly broadened components of the NLP packet. Consequently, the weaker and more fluctuating sub-pulses are filtered out, leading to a reduction in the overall energy output compared to the NPR configuration.

The right column of Fig. 4 shows the evolution of the packet duration as a function of pump power for the three output coupling percentages. In contrast with the evolution of energy, there is no clear differences between the different mode-locking mechanisms. Indeed, while the pump power is increased, the pulse duration $\Delta\tau$ increases on average in the range of 1 to 6 ns. Note that large variations of $\Delta\tau$ occurs as a direct consequence of the stochastic nature of the noise-like pulses [22]. Parameter $\Delta\tau$ is calculated by averaging over hundreds of packets.

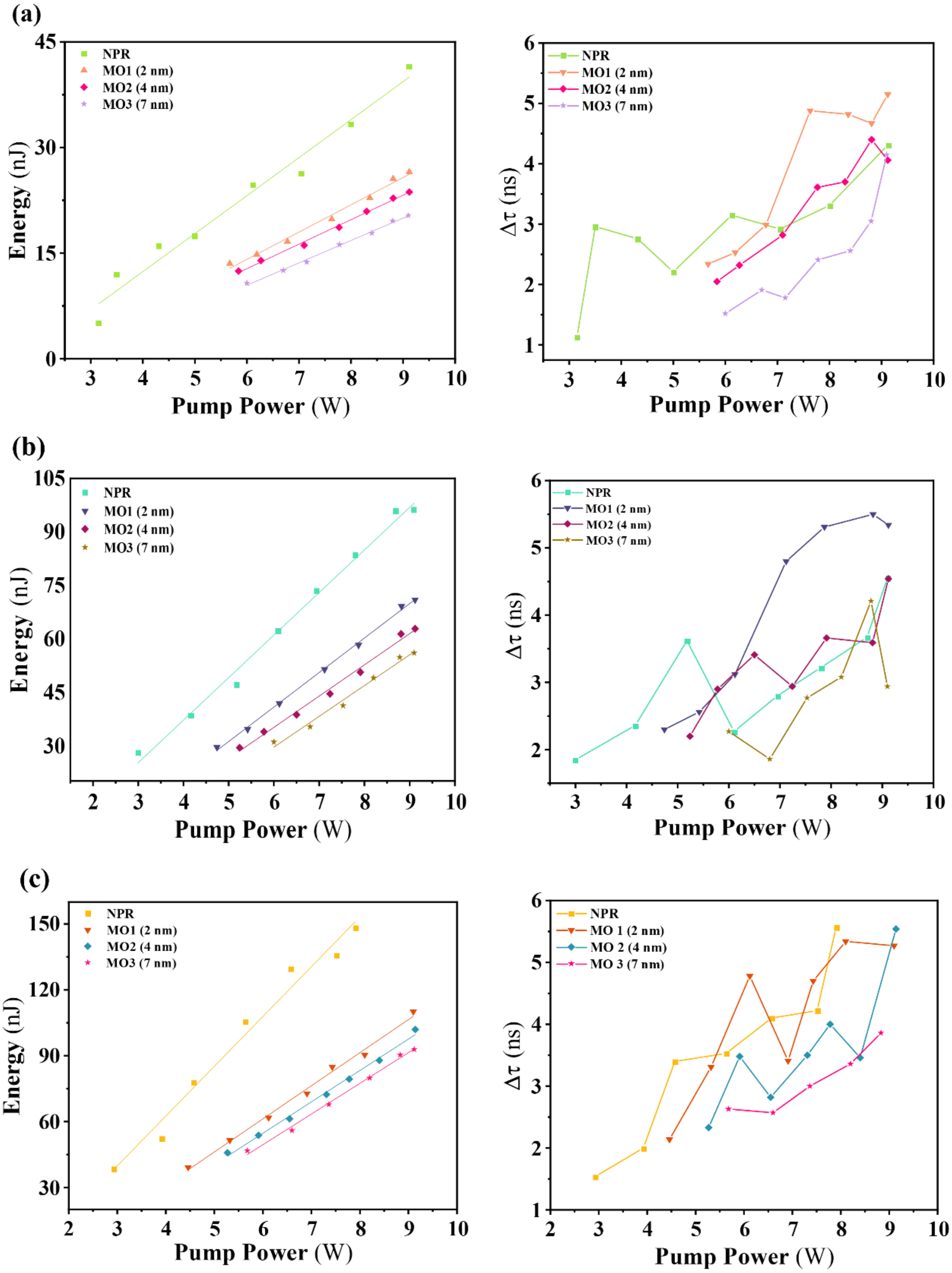

**Fig. 4**. Comparative measurements at different output coupling ratios. Left: energy versus pump power, the straight lines are linear fit, Right: NLP duration versus pump power. The output coupling coefficient $T$ is (a) 20%, (b) 50% and (c) 80%. With (NPR) denoting the conventional mode-locking configuration, while (MO1, MO2, MO3) corresponds to the Mamyshev configurations with filter separations $\delta\lambda = 2$ nm, 4 nm and 7 nm, respectively.

NLPs are also characterized by their large and nearly flat optical spectra. Experimental spectra, obtained for a fixed pumping power of 7 W and for each output coupling coefficient $T = 20\%; 50\%, 80\%$, are given in Fig. 5 a), b) and c), respectively. They show that the spectral bandwidth slightly increases when the filter separation $\delta\lambda$ increases as expected due to enhanced SPM occurring in the Mamyshev oscillator. In addition, results show that the output coupling coefficient has a notable influence on the spectral width which decreases when $T$ increases. This could be physically explained by the fact that while $T$ increases, the intra-cavity energy of the ultrashort pulses decreases leading to a reduction of the spectral broadening resulting from the Kerr effect. Let us finally mention that the spectral bandwidth undergoes unsignificant variations versus the pumping power, as commonly observed in NLP regime [6].

The best performances in terms of energy are obtained for high output coupling ($T = 80\%$) with the NPR technique which leads to 150 nJ versus 100 nJ for the MO. This is attributed to the strong filtering of low energetic ultrashort pulses in the MO case. If we look at the spectral bandwidth, the Mamyshev oscillator has better performances than NPR configuration (the spectral bandwidth at $-10$ dB level goes from 122 nm (NPR) to 166 nm (MO). This is a direct consequence of the Mamyshev regenerator which favors high energetic short pulses compared to the NPR case.

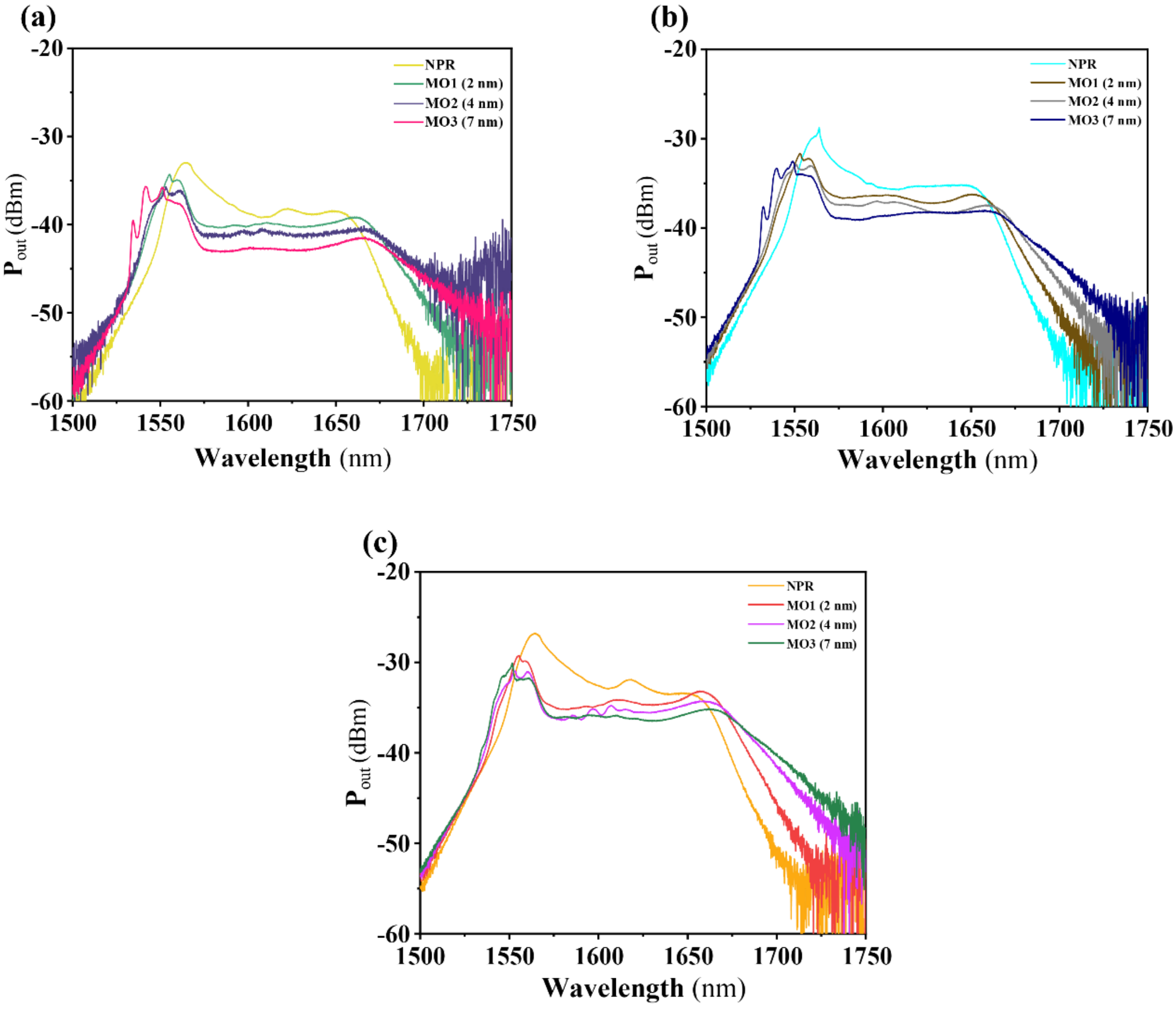

**Fig. 5.** Optical spectra obtained for a pumping power of 7 W. (a) $T = 20\%$, (b) $T = 50\%$ and (c) $T = 80\%$. (NPR) stands for the NPR mode-locking configuration, while (MO1, MO2, MO3) corresponds to the Mamyshev configuration with filter separations $\delta\lambda = 2$ nm, 4 nm and 7 nm, respectively.

## 4. Normal dispersion regime

In this section, we extend the previous investigations to the normal dispersion regime. For that we modify the experimental setup to operate in the normal dispersion regime by inserting segments of dispersion shifted fibers (DSF) in each MR arm. More precisely, two pieces of 3.5 m long of DSF ($\beta_2 = 0.16$ ps$^2$/m) were inserted just before the amplifiers, the resulting total cavity length was then $L = 52$ m. The total average dispersion of the cavity in this configuration was estimated to be approximately 0.18 ps$^2$. The output signal was extracted through a fiber coupler, with an output coupling coefficient $T = 20\%$ or $T = 50\%$. Unfortunately, unlike in the anomalous dispersion regime, higher output coupling ratios (80%) could not be achieved, as even a slight filters separation resulted in the disappearance of the NLP regime.

As in the anomalous dispersion case, for each set of parameters (output coupling ratio $T$, filters separation $\delta\lambda$) we investigate the characteristics of the NLP emission. However, in contrast to the former case, some limitations occur. The first one is related to the output coupling coefficient which is limited to 50% as mentioned previously. The second limitation concerns the maximum filter separation which is reduced to 5 nm instead of 7 nm in the anomalous case. Therefore, for the Mamyshev oscillator configuration we will consider the set of parameters $T = 20\%, T = 50\%$ and $\delta\lambda = 2$ nm, $\delta\lambda = 5$ nm. The experimental results are presented in Fig. 6, where (a) and (b) correspond to T = 20% and T = 50%, respectively. The left and right columns give the evolution of the pulse energy and duration as a function of the pumping power, respectively. The results are very similar to those obtained in the anomalous dispersion regime. The NPR systematically leads to higher energies than MO. The highest energies are obtained at the largest output coupling ratios, regardless of the mode-locking mechanism, reaching 100 nJ in the NPR regime and 75 nJ in the MO regime at a pump power of 9 W. The pulse duration increases versus the pumping power but there are no significant differences between the NPR and MO configurations.

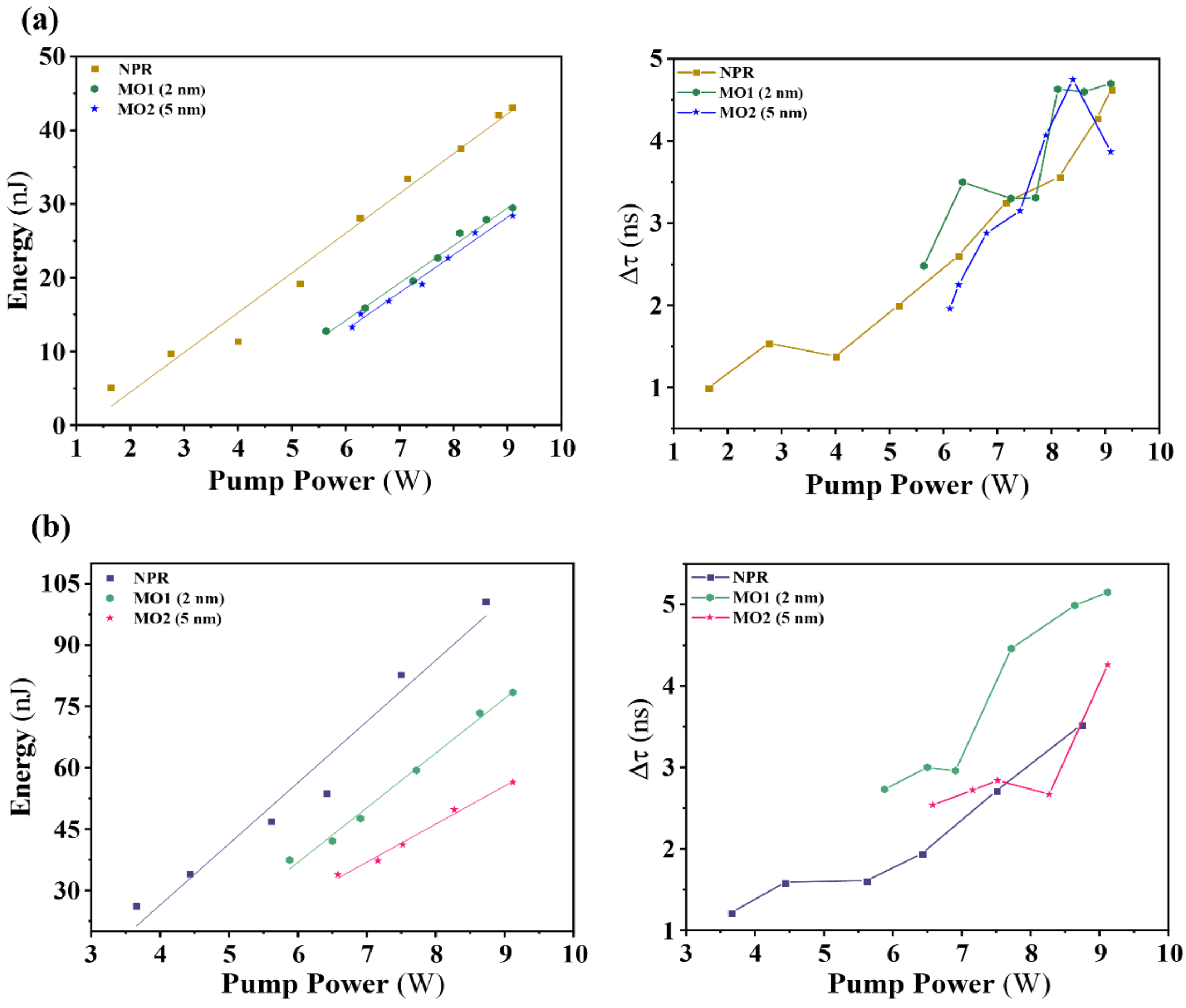

**Fig. 6**. Comparative measurements at different output coupling ratios. Left: energy versus pump power, the straight lines are linear fit, Right: NLP duration versus pump power. The output coupling coefficient $T$ is (a) 20%, (b) 50%. With (NPR) denoting the conventional mode-locking configuration, while (MO1, MO2) correspond to the Mamyshev configurations with filter separations $\delta\lambda = 2$ nm and 5 nm, respectively.

Fig. 7 shows the optical spectrum for the different optical configurations. Fig 7(a) stands for measurements obtained with an output coupling coefficient of 20% and Fig 7(b) for an output coupling coefficient of 50%. In contrast with the anomalous dispersion regime, there is more important difference between the NPR and the MO cases. Indeed, the spectral width is significantly broader in the MO configuration compared to the NPR, especially for low output coupling coefficient (Fig 7(a)). In the conventional mode-locking through NPR, such a significant difference between NLP emission in anomalous and normal dispersion regime has been previously reported [5]. The spectral bandwidth versus the output coupling exhibits a behavior similar to that observed in the anomalous dispersion regime for the MO case. The same physical explanation can therefore be given. Surprisingly, in the NPR mode-locking, the spectral bandwidth increases when $T$ increases, which is the opposite of what happens in all the other cases in the anomalous dispersion regime. We don't yet have a physical interpretation of such behavior. Finally, as in the anomalous dispersion case, the spectral width is systematically greater in the MO configuration than in the NPR geometry.

In summary, the conditions of optimization are identical to those of the anomalous dispersion case. The highest energy is obtained for high output coupling ($T = 50\%$) with the NPR technique which leads to about 100 nJ versus 75 nJ for the MO. The spectral bandwidth is larger for the MO oscillator (approximately 165 nm).

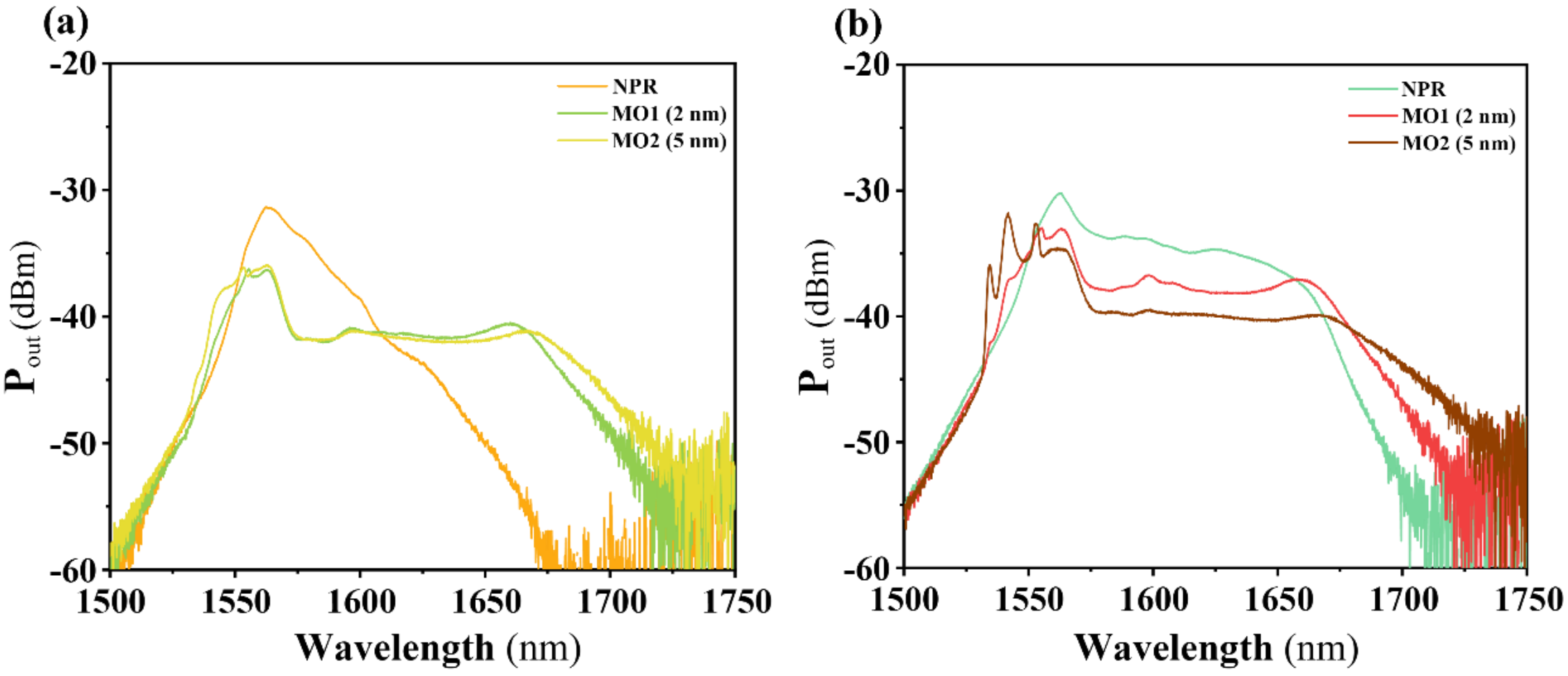

**Fig. 7.** Optical spectra obtained for a pumping power of 7 W. (a) $T = 20\%$ and (b) $T = 50\%$. (ML) stands for the NPR mode-locking configuration, while (MO1, MO2) corresponds to the Mamyshev configuration with filter separations $\delta\lambda = 2$ nm and 5 nm, respectively.

## 5. Conclusion

In summary, we have performed an extensive comparative study of noise-like pulse emission from Mamyshev oscillator and nonlinear-polarization rotation-based mode-locking in an erbium-doped fiber laser. Both anomalous and normal dispersion cases have been considered together with different output coupling coefficients. Our results demonstrate that dispersion has low qualitative influence on the comparative performances achievable from NPR mode-locking and MO. Indeed, the NPR configuration allows to obtain higher energetic NLPs than MO, independently of both the dispersion regime and the value of the output coupling coefficient. Larger spectral width is achieved from the MO compared to NPR. These results have been attributed to the strong filtering effect occurring in the MO which rejects the low energetic short pulses. Therefore, the energy of an NLP in the MO reduces compared to NPR, while the self-phase modulation increases leading to a larger spectral bandwidth. As usual in fiber laser, the highest output coupling ratio allows to extract the most energetic NLPs from the cavity independently of the model-locking mechanism and of the dispersion regime.